\documentclass[a4paper]{jpconf}
\usepackage{graphicx}
\usepackage[english]{babel}
\usepackage{amsmath,amssymb}
\usepackage{bm}

\usepackage{url}

\usepackage{xspace}

\def\araa{ARA\&A}
\def\apj{ApJ}
\def\apjl{ApJ}
\def\apjs{ApJS}

\def\aap{A\&A}

\def\aaps{A\&AS}

\def\mnras{MNRAS}

\def\pasp{PASP}

\def\nat{Nature}

\def\rmxaa{RMxAA}

\newcommand{\maihem}{\textsc{maihem}\xspace}
\newcommand{\flash}{\textsc{flash}\xspace}
\newcommand{\cloudy}{\textsc{cloudy}\xspace}

\newcommand{\ionic}[2]{#1$\,${\scshape{#2}}\xspace}%    % ion, i.e., C II = \ionic{C}{ii}

\begin{document}
\title{Non-equilibrium Photoionization and Hydrodynamic Simulations of Starburst-driven Outflows}

\author{A Danehkar, M S Oey and W J Gray}

\address{Department of Astronomy, University of Michigan, 1085 S. University Ave, 
Ann Arbor, MI 48109, USA}

\ead{danehkar@umich.edu}

\begin{abstract}
Starburst-driven galactic outflows in star-forming galaxies have been observed to contain complex thermal structures and emission line features that are difficult to explain by adiabatic fluid models and plasmas in photoionization equilibrium (PIE) and collisional ionization equilibrium (CIE). We previously performed hydrodynamic simulations of starburst-driven outflows using the \maihem module for non-equilibrium atomic chemistry and radiative cooling functions in the hydrodynamics code \flash, and calculated emission lines in combined CIE and PIE conditions. In the present study, we consider time-dependent non-equilibrium ionization (NEI) states produced by our \maihem simulations. Through extensive \cloudy calculations made with the NEI states from our hydrodynamic simulations, we predict the UV and optical emission line profiles for starburst-driven outflows in time-evolving non-equilibrium photoionization conditions. Our hydrodynamic results demonstrate applications of non-equilibrium radiative cooling for \ionic{H}{ii} regions in starburst galaxies hosting cool outflows.
\end{abstract}

\section{Introduction}

Starburst-driven galactic-scale outflows are typically seen in star-forming galaxies \cite{Heckman1990,Lehnert1995,Dahlem1997,Rupke2002,Martin2005,Veilleux2005}, which
have complex multi-temperature structures ranging from cold ($100$--1000\,K) in radio and IR \cite{Ott2005,Weis2005} 
to warm ($\sim 10^4$\,K) in optical \cite{Izotov1999,James2009}, and hot ($ \sim 10^{7}$\,K) in X-ray observations \cite{Strickland1997,Ott2005a}. Hydrodynamic properties of starburst-driven outflows have approximately been modeled with  adiabatic fluid models \cite{Weaver1977,Chevalier1985,Canto2000}.
The adiabatic solution of a free expanding wind can produce a \textit{hot bubble} and a \textit{shell} for an outflow model surrounded by the ambient medium \cite{Castor1975}. Moreover, the adiabatic solution indicates that density and temperature of the free expanding wind decrease with radius $r$ as $\rho \varpropto r^{-2}$ and $T \varpropto r^{-4/3}$ \cite{Chevalier1985}. Thus, a starburst-driven outflow could have four regions: freely expanding wind, hot bubble, shell, and ambient medium \cite{Weaver1977} (see Figure~\ref{outflow}).

However, the adiabatic assumption seems to be inconsistent with strong cooling and suppressed outflows seen in several star-forming galaxies \cite{Smith2006,Turner2017,Oey2017}. Starburst-driven outflows have also been investigated using fluid models with radiative cooling functions \cite{Silich2004,Tenorio-Tagle2005}, implying that cooling depends on the metallicity, mass-loading, and wind terminal velocity. 
More recently, the occurrence of strong radiative cooling has been investigated using radiative cooling functions in non-equilibrium conditions \cite{Gray2019}, which depicted the dependence of cooling on the metallicity, mass-loading, and wind velocity in agreement with the previous numerical results derived from semianalytic models \cite{Silich2004}. Moreover, a large grid of hydrodynamic simulations indicate that radiative cooling is stronger with higher mass-loading rate, and lower wind terminal velocity \cite{Danehkar2021}.
However, emission lines predicted by this large grid of models were calculated using photoionization equilibrium (PIE) and collisional ionization equilibrium (CIE) that do not allow us to clearly identify the parameter space with strong radiative cooling \cite{Danehkar2021}.

It has been found that emission lines calculated by non-equilibrium ionization (NEI) could have a large departure from CIE \cite{Gnat2007,Vasiliev2011,Oppenheimer2013} in plasma regions ($<10^6$\,K) being in the transition from pure CIE to PIE \cite{Vasiliev2011}. Previously, cooling functions and NEI chemistry networks were included in the package \maihem (Models of Agitated and Illuminated Hindering and Emitting Media) \cite{Gray2015,Gray2016,Gray2019}. Some non-equilibrium ionization models built by \maihem showed the enhancements of \ionic{O}{vi} and \ionic{C}{iv} in non-equilibrium conditions \cite{Gray2019a}. 

Previoulsy, in Ref.~\cite{Danehkar2021}, we obtained emission lines predicted by combined CIE+PIE models of \maihem hydrodynamic simulations for different outflow parameters. In this recent work \cite{Danehkar2021b}, we calculate emission lines for combined NEI+PIE model using NEI states produced by our \maihem  hydrodynamic simulations. 

\section{Numerical Methods}
\label{numeric}

\begin{figure}
\centering
\includegraphics[width=0.6\textwidth, trim = 0 0 0 0, clip, angle=270]{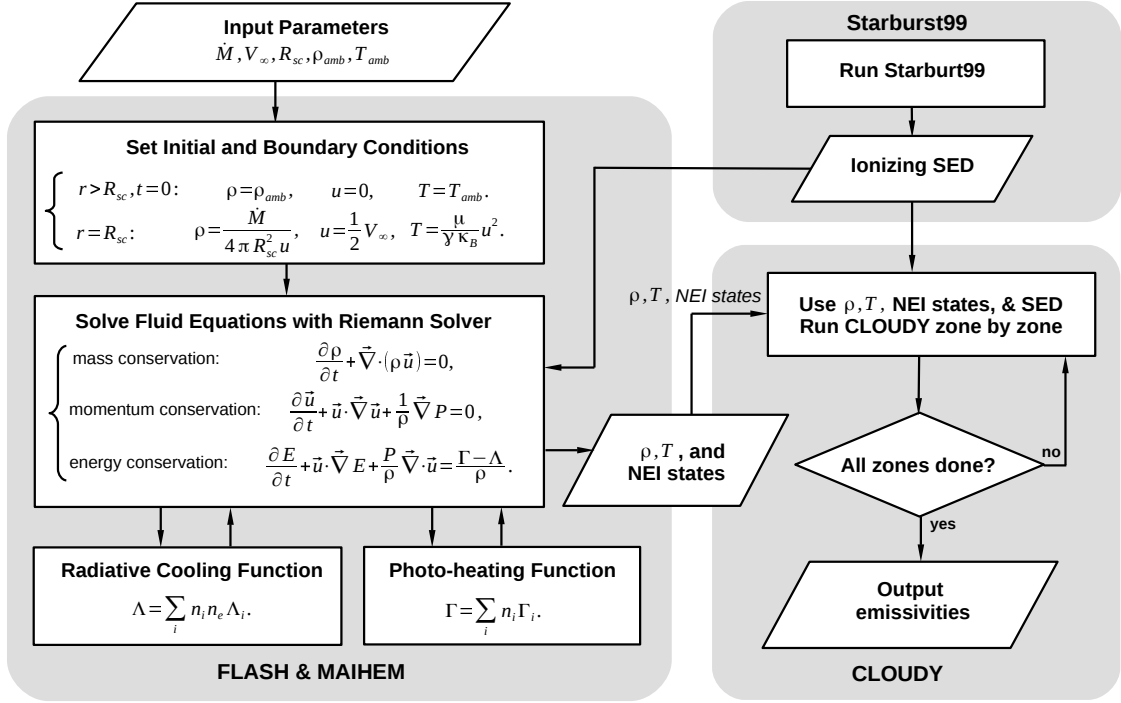}
\caption{Schematic view of the connections between \flash hydrodynamic simulations, radiative cooling and heating functions in \maihem used to produce NEI states, ionizing SED model made by Starburst99, and non-equilibrium photoionization calculations conducted by \cloudy zone by zone using the Starburst99 SED, physical conditions ($\rho$ and $T$), and NEI states generated by non-equilibrium atomic chemistry functions in \maihem.}
\label{schematic}
\end{figure}

We consider a galactic-scale outflow driven by starburst feedback from a spherically symmetric super star cluster (SSC) characterized with the SSC radius  $R_{\rm sc}$, mass deposition rate $\dot{M}$, wind terminal velocity $V_{\infty}$, and stellar ionizing fields parameterized by ionizing luminosity $L_{\rm ion}$ and spectral energy distribution (SED) surrounded by ambient medium with density $\rho_{\rm amb}$ and temperature $T_{\rm amb}$. To numerically simulate starburst-driven outflows with various parameters, we model the fluid equations with appropriate initial and boundary conditions (\S\,\ref{numeric:hydrodynamic}), ionizing radiation fields produced by SSCs (\S\,\ref{numeric:sed}), radiative thermal processes within outflows (\S\,\ref{numeric:cooling}), and non-equilibrium photoionization processes (\S\,\ref{numeric:photoionization}), as illustrated in Figure~\ref{schematic}.

\subsection{Hydrodynamic Simulations}
\label{numeric:hydrodynamic}

We use the MUSCL method \cite{vanLeer1979} and hybrid Riemann solver \cite{Toro1994} in the hydrodynamics code \flash \cite{Fryxell2000} to solve the fluid equations for mass conservation, momentum conservation, and energy conservation coupled with radiative cooling and heating functions of the NEI chemistry package \maihem \cite{Gray2015,Gray2016,Gray2019} that are as follows:
\begin{equation}
\frac{\partial \rho}{\partial t}+\vec{\nabla} \cdot \left( \rho\vec{u}\right) =0,  \label{eq_1}
\end{equation}%
\begin{equation}
\frac{\partial \vec{u}}{\partial t}+(\vec{u}\cdot \vec{\nabla}) \vec{u}+\frac{1}{\rho}\vec{\nabla} P=0,  \label{eq_2}
\end{equation}%
\begin{equation}
\frac{\partial {E}}{\partial t} + \vec{u} \cdot \vec{\nabla} E +\frac{P}{\rho} \vec{\nabla} \cdot \vec{u} =\frac{\Gamma-\Lambda}{\rho} ,  \label{eq_3}
\end{equation}%
where $\rho$, $u=\vert\vec{u}\vert$, $P=\left( \gamma -1\right) \rho\epsilon$, and $E=\epsilon+\frac{1}{2}u^{2}$ is the fluid density, velocity, and thermal pressure, and total energy per mass, respectively, $\epsilon$ the internal energy per mass, $\gamma=5/3$ the heat capacity ratio, and $\Lambda$ and $\Gamma$ the radiative cooling and heating rates per volume, respectively.

\begin{figure}
\centering
\includegraphics[width=0.45\textwidth, trim = 0 0 0 0, clip, angle=0]{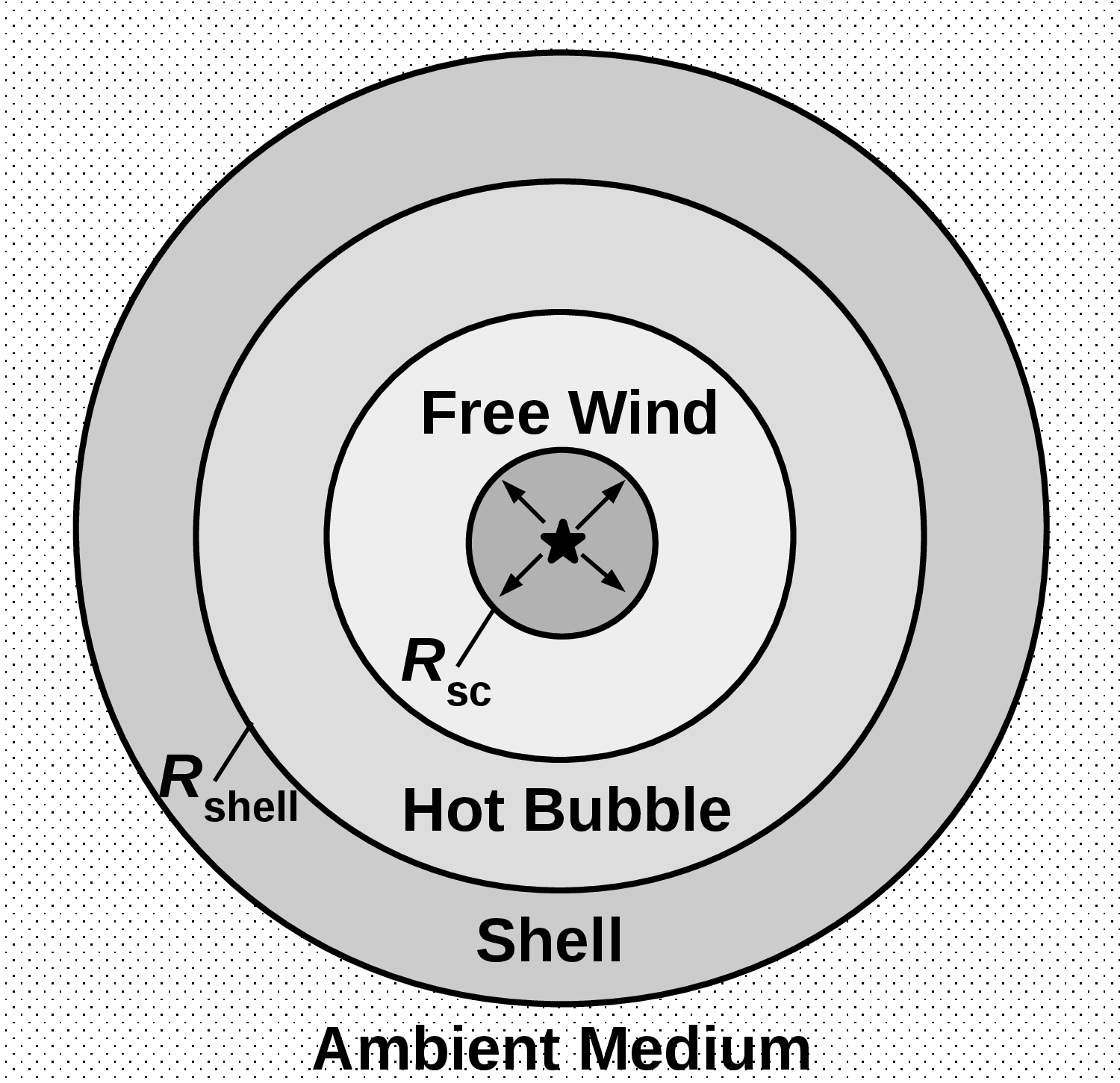}~~~~%
\includegraphics[width=0.45\textwidth, trim = 0 0 0 0, clip, angle=0]{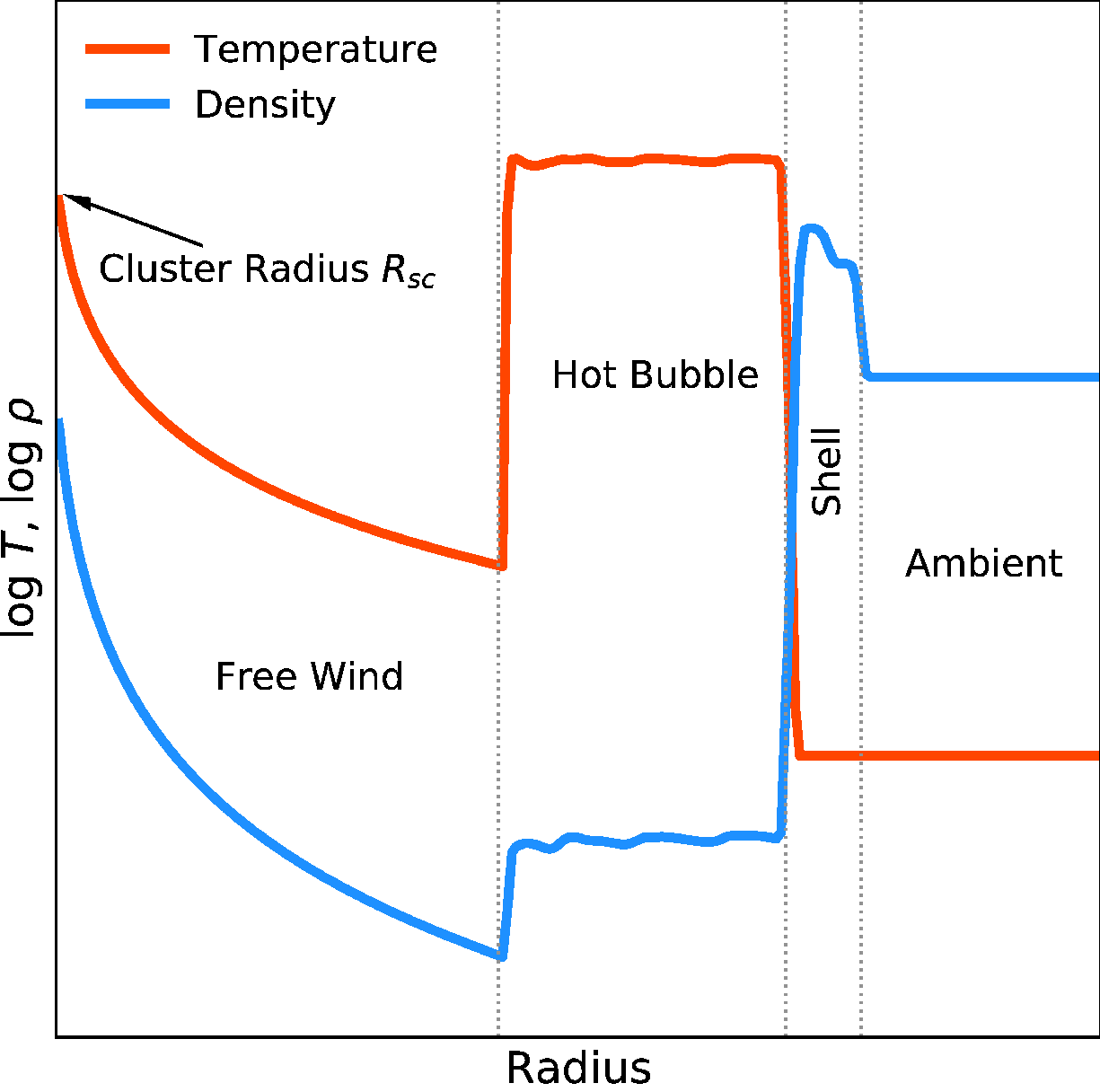}
\caption{Schematic view of the starburst-driven outflows (left panel) with the temperature profile (plotted by red line; right panel) and density profile (blue line), showing the different regions (separated by dotted lines)
described by \cite{Weaver1977}: free wind, hot bubble, shell, and ambient medium.
}
\label{outflow}
\end{figure}

To produce a starburst-driven outflow defined by $\dot{M}$ and $V_{\infty}$ at the cluster boundary ($r=R_{\rm sc}$) surrounded by the ambient medium, we use the analytic solutions of the steady-state outflow model in one-dimensional spherical coordinates  
\cite{Chevalier1985,Canto2000,Silich2004}, and set density $\rho$, velocity $u$, and temperature $T$ at $r=R_{\rm sc}$ 
with user-defined non-penetrating boundary conditions as follows:
\begin{equation}%
\begin{array}
[c]{ccc}%
\rho \Big|_{r=R_{\rm sc}} = \dfrac{ \dot{M}}{ 4 \pi  R_{\rm sc}^2 u},~~~~~
& u\Big|_{r=R_{\rm sc}} = \dfrac{1}{2} V_{\infty} , ~~~~~
T\Big|_{r=R_{\rm sc}} =   \dfrac{\mu}{\gamma k_{\rm B}} u^2 ,
\end{array}
\end{equation}
where $\mu$ is the mean mass per particle and $k_{\rm B}$ the Boltzmann constant.

We also configure the initial values of the density, temperature, and velocity at $t=0$ for the stationary ambient medium ($r>R_{\rm sc}$) with $\rho_{\rm amb}$ and $T_{\rm amb}$ surrounding the SSC:
\begin{equation}%
\begin{array}
[c]{ccc}%
\rho \Big|_{r>R_{\rm sc},\,t=0} = \rho_{\rm amb},~~~~~
& u\Big|_{r>R_{\rm sc},\,t=0} = 0 , ~~~~~
T\Big|_{r>R_{\rm sc},\,t=0} =   T_{\rm amb}. 
\end{array}
\end{equation}
The outer boundary at $r=250$ pc sets zero-gradient boundary conditions that allow the outflow to leave the simulation domain. We note for an ideal gas $P= \rho k_{\rm B} T /\mu$. 

We solve Eqs. (\ref{eq_1})--(\ref{eq_3}) using the unsplit hydrodynamic solver \cite{Lee2009,Lee2009a,Lee2013} included in \flash  with the 2nd-order MUSCL--Hancock scheme \cite{vanLeer1979}. To have stable solutions in regions with strong shocks, we utilize a hybrid Riemann solver \cite{Toro1994} that includes Roe \cite{Roe1981} and HLLE solvers \cite{Einfeldt1988,Einfeldt1991}. Simulations are performed on a grid with $N_{x}=512$ blocks and maximum refinement levels of two, while the refinement criterion \cite{Loehner1987} is set to follow the fluid variables $\rho$, $T$, $P$, and $u$. This provides a maximum resolution of 0.244\,pc.

\subsection{Ionizing Radiation Source Modeling}
\label{numeric:sed}

Radiation fields produced by stars within the SSC result in photoionization and radiative processes. 
To model the ionizing radiation source for our hydrodynamic simulations and non-equilibrium photoionization calculations, we generate a SED using the stellar population synthesis program Starburst99 \cite{Leitherer1999,Levesque2012,Leitherer2014} for a given total stellar mass $M_{\star}=2\times10^6$\,M$_{\odot}$ corresponding  to a mass-loading rate of $\dot{M} = 10^{-2}$\,M$_{\odot}$\,yr$^{-1}$ at solar metallicity and age of 1 Myr. For our Starburst99 model, we employ an IMF with slope $\alpha= 2.35$ in the mass range 0.5--150 M$_{\odot}$, the 
Geneva rotational stellar evolution models \cite{Ekstroem2012,Georgy2012,Georgy2013}, and the extended Pauldrach/Hillier stellar atmosphere \cite{Hillier1998,Pauldrach2001}. The ionizing luminosity and SED spectrum calculated by Starburst99 are supplied to photo-heating functions of \maihem and non-equilibrium photoionization calculations performed by \cloudy. To make Starburst99 outputs usable for \cloudy, we compile a binary format ``mod'' file from the Starburst99 ``spectrum'' output file using the \cloudy command \textsf{compile star}. Moreover, we determine the ionizing luminosity ($L_{\rm ion}$) from the total luminosity and H$^{+}$ fraction at age 1\,Myr read from the Starburst99 ``quanta'' output file.

\subsection{Radiative Cooling and Heating Processes}
\label{numeric:cooling}

To have radiative thermal effects in hydrodynamic simulations, it is necessary to calculate radiative cooling and photo-heating rates. The cooling routine in the \maihem package implemented by Ref.\,\cite{Gray2015} computes the radiative cooling rate $\Lambda$ using the ion-by-ion cooling efficiencies $\Lambda_i$ for different species \cite{Gnat2012} that depend on the metallicity $Z$ and temperature $T$ produced by hydrodynamic simulations. The heating routine in \maihem added by Ref.\,\cite{Gray2019} calculates the photo-heating cooling rate $\Gamma$  using the photoionization cross section \cite{Verner1995,Verner1996}.

The radiative cooling $\Lambda$ and heating rates $\Gamma$ are calculated using the ion number density $n_i$, and the electron number density $n_e$:
\begin{equation}%
\begin{array}
[c]{cc}%
\Lambda  = \displaystyle\sum_{i}^{}  n_i n_e \Lambda_i,~~~~~
\Gamma = \displaystyle\sum_{i}^{}  n_i  \Gamma_i, 
\end{array}\label{eq:7}%
\end{equation}
where $\Lambda_i$ are the cooling efficiencies from \cite{Gnat2012},  $\Gamma_i$ the heating efficiencies estimated using the SED spectrum from Starburst99 and photoionization cross sections from \cite{Verner1995,Verner1996}. 

\begin{figure}
\centering
\includegraphics[width=0.33\textwidth, trim = 0 0 0 0, clip, angle=0]{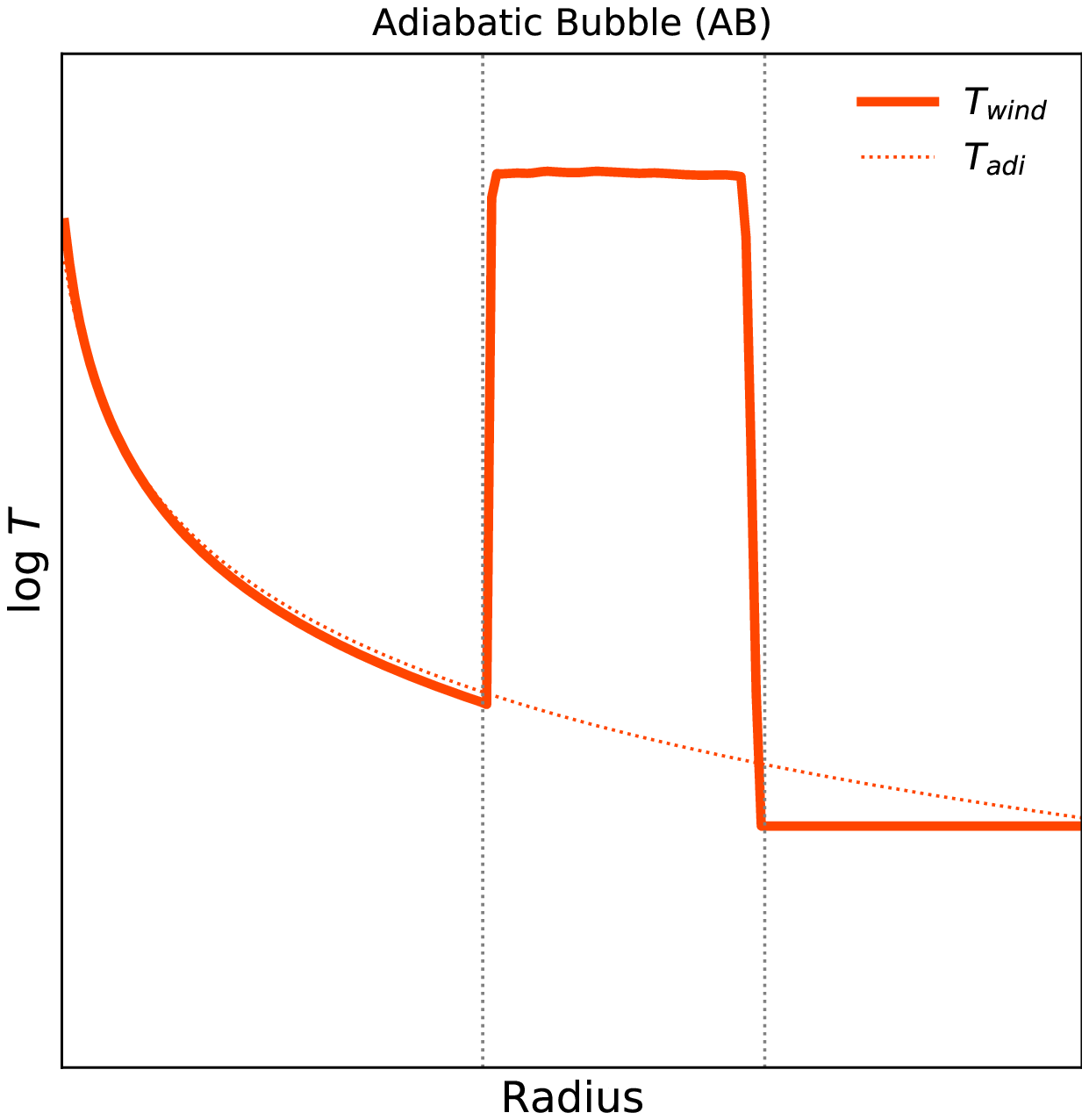}%
\includegraphics[width=0.33\textwidth, trim = 0 0 0 0, clip, angle=0]{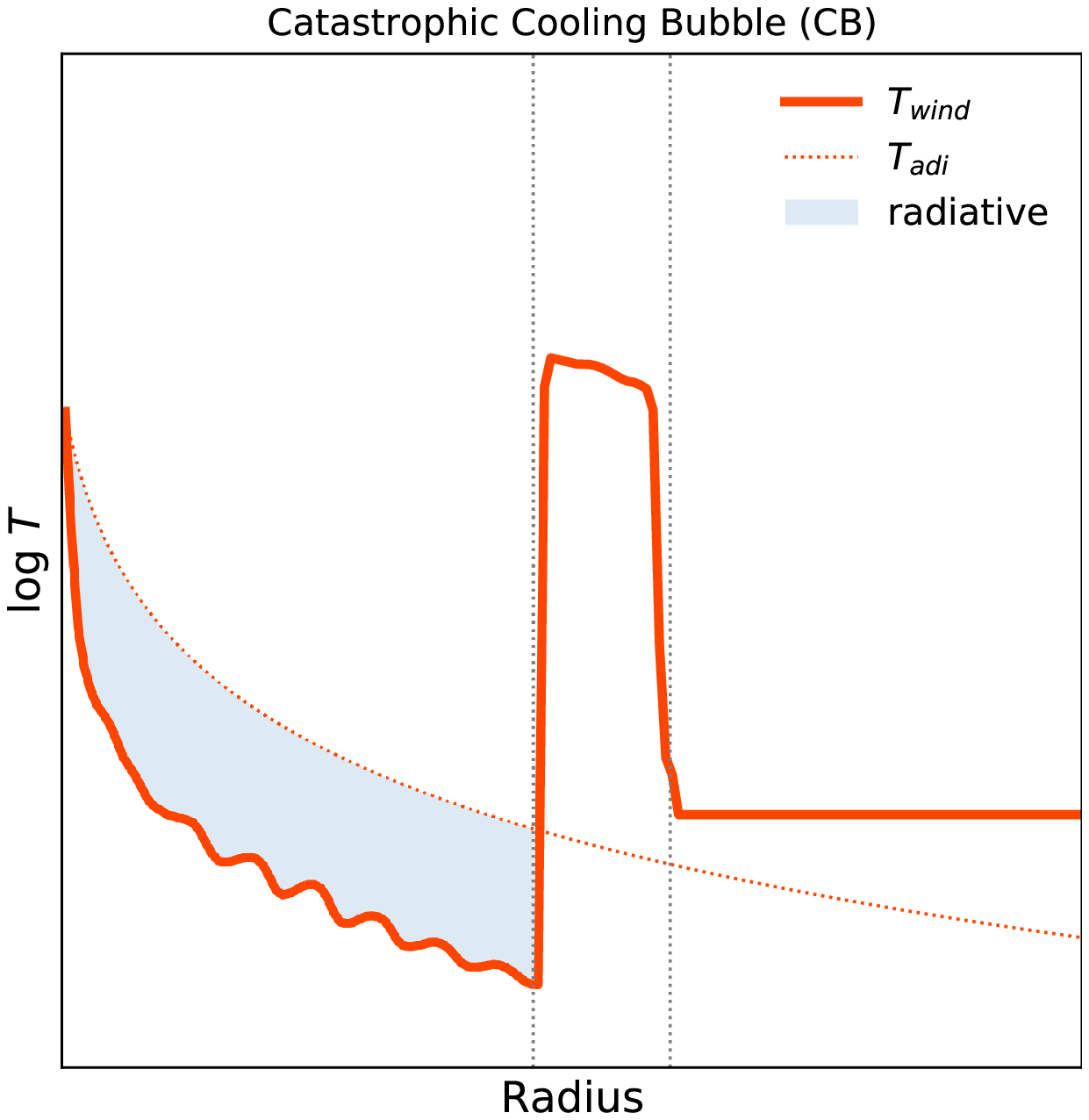}%
\includegraphics[width=0.33\textwidth, trim = 0 0 0 0, clip, angle=0]{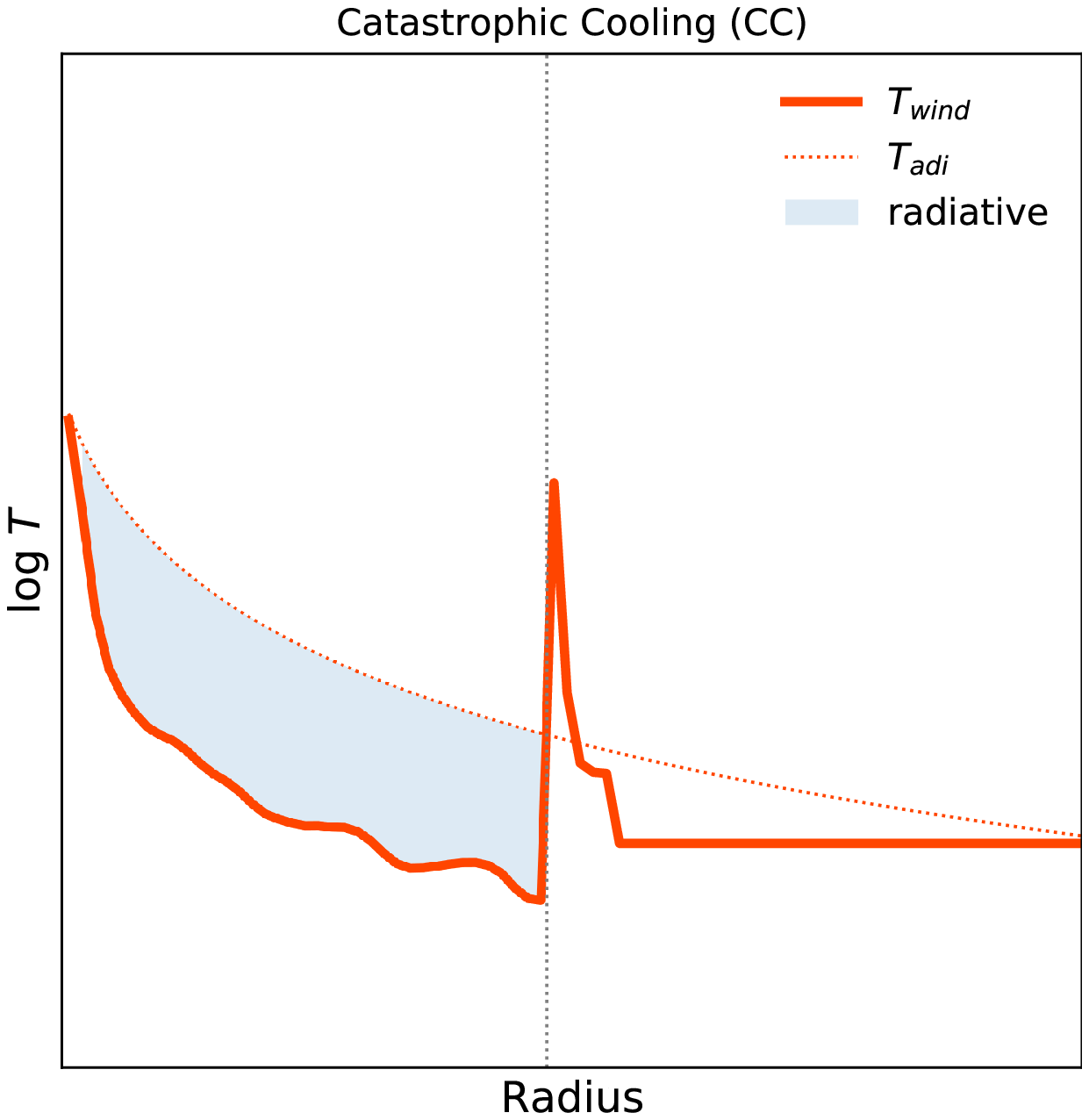}
\caption{Various wind modes based on wind temperature profiles: adiabatic bubble (AB), 
catastrophic cooling bubble (CB), and catastrophic cooling (CC), described in \cite{Danehkar2021}.
Temperature profiles and their predicted adiabatic solutions are shown by solid and dotted red lines, respectively. 
Radiative cooling in free winds contributes to departures from the expected adiabatic solutions. 
The boundaries of the bubble are also shown by dotted gray lines.
}
\label{flash}
\end{figure}

\begin{figure}
\centering
\includegraphics[width=0.45\textwidth, trim = 0 0 0 0, clip, angle=270]{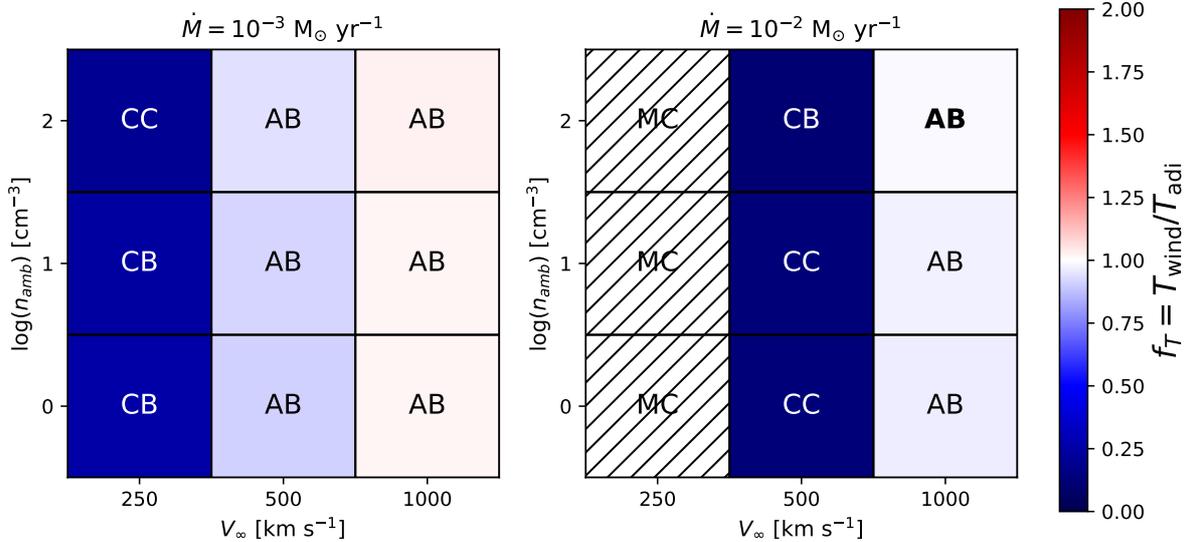}
\caption{The wind temperature with radiative cooling relative to the predicted adiabatic temperature, $f_T \equiv T_{\rm wind} / T_{\rm adi}$, in the parameter space of the wind terminal velocity $V_{\infty}=250$, 500, and 1000 km\,s$^{-1}$, 
ambient number density $n_{\rm amb}=1$, 10, and 100 \,cm$^{-3}$, and  mass-loading rate $\dot{M}= 10^{-3}$ and $10^{-2}$\,M$_{\odot}$\,yr$^{-1}$ for the models with the solar metallicity, cluster radius $R_{\rm sc} =1$\,pc, total stellar mass $M_{\star}=2 \times 10^6$\,M$_{\odot}$, and age $t=1$ Myr.
Outflow models are classified according to radiative/adiabatic cooling free winds and presence/absence of bubbles in their temperature profiles, namely adiabatic bubble (AB), catastrophic cooling (CC),  catastrophic cooling bubble (CB), and momentum-conserving (MC), see \cite{Danehkar2021}.  
An optically-thick AB model is shown by the bold font. 
}
\label{wind:mode}
\end{figure}

Moreover, the \maihem package as implemented by \cite{Gray2015} also contains the atomic chemistry functions that perform non-equilibrium ionization, collisional ionization and recombination processes using collisional ionization rates \cite{Voronov1997}, recombination rates \cite{Badnell2006}, and dielectronic recombination rates \cite{Gray2015}. Thus, \maihem supplies \flash hydrodynamic simulations with radiative thermal functions and non-equilibrium ionization that allow us to study time-dependent radiative effects in starburst-driven outflows (see Figures~\ref{flash} and \ref{wind:mode}; described in \S\,\ref{results}).

\subsection{Non-equilibrium Photoionization Modeling}
\label{numeric:photoionization}

Non-equilibrium photoionization models are built by the program \cloudy 
\cite{Ferland1998,Ferland2013,Ferland2017} using the density profile $\rho (r)$, temperature profile $T(r)$, and NEI states $F_{\rm i} (r)$ of outflows as a function of $r$ predicted by hydrodynamic simulations with \maihem (see Figure~\ref{cloudy}; described in \S\,\ref{results}), together with the ionizing source made by Starburst99. The procedure is fully described in \cite{Gray2019a} (non-CIE models). Previously, we also produced CIE+PIE models using $\rho (r)$, $T(r)$, and SED, but without $F_{\rm i}(r)$ \cite{Danehkar2021}, which incorporate hydrodynamic collisional ionization and photoionization processes.

Modeling non-equilibrium photoionization requires performing multiple zone-by-zone \cloudy runs on physical conditions and NEI states with the radial distance $r$ ranging from the cluster radius $R_{\rm sc}$ to the shell radius $R_{\rm shell}$. For each zone ($r$, $r+\vartriangle r$), we run one \cloudy run for its density ($\rho$) specified with the \cloudy command \textsf{hden}, temperature ($T$) through the command \textsf{constant temperature}, and ionization states ($F_{\rm i}$) using the \cloudy command \textsf{element name ionization}. Multiple \cloudy runs compute the emissivities through iterations over all the outflow zones, i.e. $r \in [R_{\rm sc}, R_{\rm shell}]$. The SED (and ionizing luminosity) made by Starburst99 is supplied to \cloudy models (via the command \textsf{table star}). As this procedure needs multiple \cloudy runs for all the zones, it is computationally expensive. This procedure continues until it reaches the shell boundary $R_{\rm shell}$. Afterwards, we apply the emissivities obtained 
from the CIE+PIE models in \cite{Danehkar2021} to isothermal zones of the shell and the ambient medium.

\section{Results}
\label{results}

Figure~\ref{outflow} (right) shows the temperature $T$ and density $\rho$ as a function of radius $r$ predicted by our \maihem simulation for an adiabatic example. It has four different regions described in \cite{Weaver1977}: (a) freely expanding wind, (b) a hot bubble, (c) a narrow dense shell, and (d) ambient medium. Ref.~\cite{Danehkar2021} classified the the temperature profile of a freely expanding wind under the adiabatic bubble (AB), catastrophic cooling bubble (CB), and catastrophic cooling (CC) wind modes. The AB mode is the classic bubble model described by \cite{Weaver1977} and shown in Figure~\ref{outflow} (right). In the CB mode, the temperature profile in the region (a) has radiative cooling, but it still has a hot bubble. In CC mode, the temperature profile of the free wind is also radiatively cooled, but it is without any hot bubble. In Figure~\ref{flash}, we plot the temperature profiles for three example models in the AB, 
CB, and CC modes with expected adiabatic temperature profiles in dotted red lines.

\begin{figure}
\centering
\includegraphics[width=0.7\textwidth, trim = 0 0 0 0, clip, angle=270]{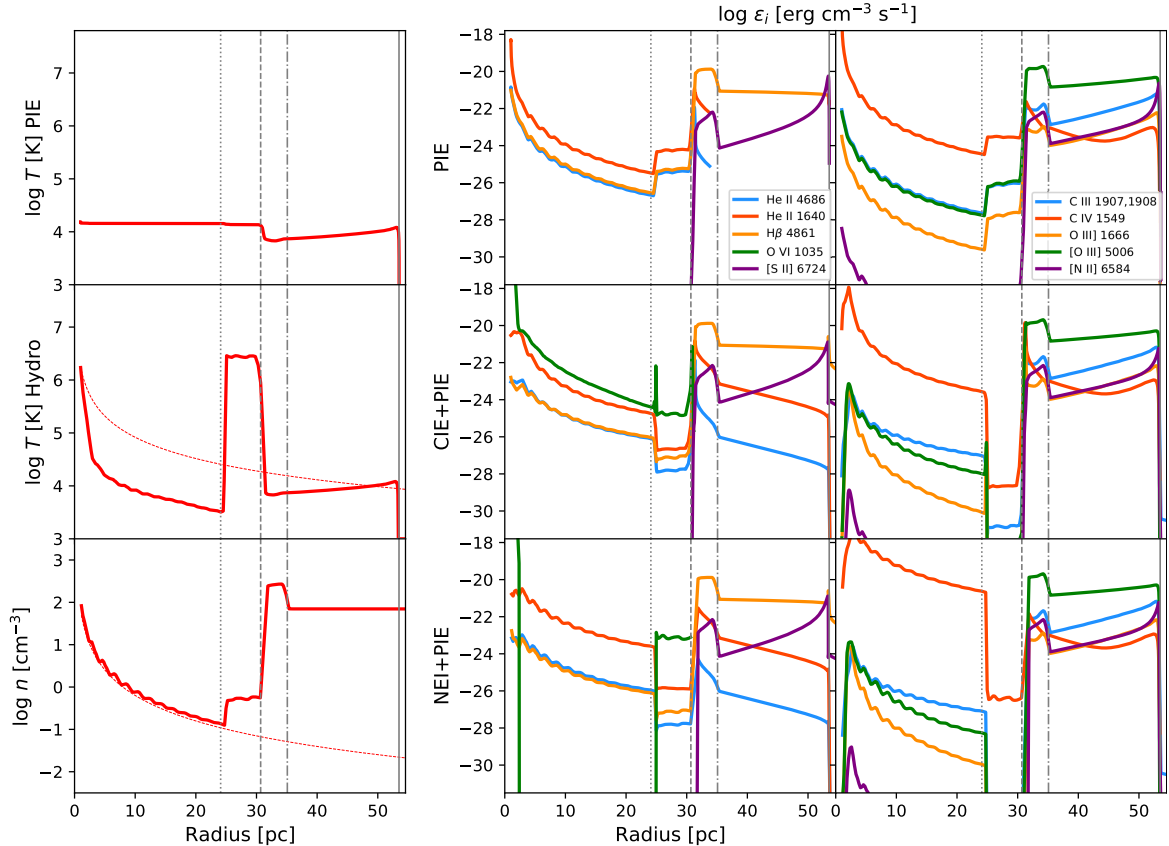}
\caption{Temperature profiles $T(r)$ produced by the pure photoionization (PIE) model and \maihem hydrodynamic simulation, and density profile $n(r)$ generated by the hydrodynamic simulation (left panels) along with the adiabatic predictions for hydrodynamic simulation (red dashed lines), 
and the volume emissivities $\epsilon_{i}(r)$ of emission lines (right panels) predicted by the pure photoionization (PIE), 
combined photoionization and collisional ionization (CIE+PIE) from \cite{Danehkar2021}, combined non-equilibrium ionization and photoionization (NEI+PIE) models from \cite{Danehkar2021b}
The boundaries of the bubble and the end of the shell are shown by dotted, dashed, and dash-dotted, respectively. 
The model parameters are: $V_{\infty}=500$\,km\,s$^{-1}$, $n_{\rm amb}=100$\,cm$^{-3}$, $\dot{M}= 10^{-2}$\,M$_{\odot}$\,yr$^{-1}$, $R_{\rm sc} =1$\,pc, $M_{\star}=2 \times 10^6$\,M$_{\odot}$, solar metallicity, and age $t=1$ Myr.
}
\label{cloudy}
\end{figure}

Figure~\ref{wind:mode} depicts the wind temperature predicted by radiative cooling over the expected adiabatic temperature, $f_T \equiv T_{\rm wind} / T_{\rm adi}$, along with the different wind modes in the parameter space of the wind terminal velocity, ambient density, and  mass-loading rate, where the cluster radius is $R_{\rm sc} =1$\,pc, total stellar mass $M_{\star}=2 \times 10^6$\,M$_{\odot}$, solar metallicity, and age $t=1$ Myr. The CC and CB modes are associated with values of $f_T$ less than 0.75, whereas $0.75<f_T <1.25$ for the AB mode. Additionally, we assign the momentum-conserving (MC) mode to those models having totally suppressed wind due to catastrophic cooling (see \cite{Danehkar2021} for details). We also display the AB mode of an optically-thick model with the bold font, whose ambient medium is not ionized. We see that higher mass-loading rates and lower wind terminal velocities lead to strong cooling effects in the free expanding wind.

Figure~\ref{cloudy} (left panels) shows the temperature profiles $T(r)$ derived from one example of our hydrodynamic simulations\footnote{Our results for all the hydrodynamic simulations and photoionization calculations are presented as an interactive figure in \cite{Danehkar2021} and hosted on this URL: \url{https://superwinds.astro.lsa.umich.edu/}} with pure photoionization (PIE) and combined hydrodynamic and photoionization thermal effects, the density profile $n(r)$ predicted by the hydrodynamic simulation. Figure~\ref{cloudy} (right panels) shows the emissivities obtained for one hydrodynamic simulation with pure photoionization (PIE), combined hydrodynamic collisional ionization and photoionization (CIE$+$PIE), and combined non-equilibrium ionization and photoionization (NEI$+$PIE). The luminosities of optical and UV emission lines are used to produce diagnostic diagrams based on volume-integrated emissivities from CIE$+$PIE models \cite{Danehkar2021} and NEI$+$PIE models \cite{Danehkar2021b}.

\section{Conclusion}

We have conducted hydrodynamic simulations of starburst-driven outflows using the \maihem module in \flash to study the formation of radiatively cooling, expanding winds, hot bubbles, and narrow dense shells (see Figure~\ref{outflow}, right) for various wind terminal velocity, ambient density, and mass-loading rate (see Figure~\ref{wind:mode}, also see \cite{Danehkar2021}). Our modeling is implemented using the same computational methods developed by \cite{Gray2019a} with boundary conditions based on the semianalytic results found by \cite{Silich2004}. Our findings confirm that radiative cooling is increased by raising mass-loading rate and reducing wind terminal velocity  \cite{Danehkar2021}.

In the previous work \cite{Danehkar2021}, we predicted emissivities for the density and temperature profiles generated by our hydrodynamic simulations in PIE and CIE+PIE conditions, while we additionally calculate emissivities in NEI+PIE conditions in this work (see Figure~\ref{cloudy}). Ionization states in PIE and CIE+PIE conditions were calculated by \cloudy (fully described in \cite{Danehkar2021}), whereas in NEI+PIE conditions we also provide \cloudy with time-dependent NEI states created by \maihem \cite{Danehkar2021b}. As found by \cite{Gray2019a}, \ionic{C}{iv} emissivity predicted by NEI+PIE model is stronger than those made by PIE and CIE+PIE models (see Figure~\ref{cloudy}, right), especially for strong cooling.

Our non-equilibrium photoionization models are conducted using a fixed typical cluster radius (1\,pc) and an ionizing source associated with a fixed total stellar mass ($M_{\star}=2 \times 10^6$\,M$_{\odot}$) and a final age of 1\,Myr. We are carrying out further  hydrodynamic simulations to explore other parameters contributing to radiative cooling in starburst-driven outflows such as cluster radius,  age, and time-dependent ionizing source.

\ack{}
\flash\ was developed in part by DOE NNSA-ASC OASCR Flash Center at U.\,Chicago.

\section*{References}
%\bibliography{iopart-num}
%\bibliography{references}{}
%\bibliographystyle{iopart-num-v2}
\providecommand{\newblock}{}

\end{document}